\documentclass[epj,english]{svjour}
\usepackage[T1]{fontenc}
\usepackage{ae,aecompl}
\usepackage[latin1]{inputenc}
\usepackage{graphicx}
\usepackage{amssymb}

\makeatletter

\providecommand{\LyX}{L\kern-.1667em\lower.25em\hbox{Y}\kern-.125emX\@}
\usepackage{graphicx}
\usepackage{graphicx}
\usepackage{dcolumn}
\usepackage{bm}

\usepackage{babel}
\makeatother
\begin{document}

%\preprint{1.4}

\title{Atomic motion in tilted optical lattices}

\author{Quentin Thommen \and
Jean Claude Garreau \and
Véronique Zehnlé}

\institute{Laboratoire de Physique des Lasers, Atomes et Mol\'{e}cules\\
 Centre d'Etudes et de Recherches Laser et Applications, Universit\'{e}
des Sciences et Technologies de Lille, \\
 F-59655 Villeneuve d'Ascq Cedex, France
\thanks{http://www.phlam.univ-lille1.fr}}

\date{\today{}}

\abstract{
This paper presents a formalism describing the dynamics of a quantum
particle in a one-dimensional, time-dependent, tilted lattice. The
formalism uses the Wannier-Stark states, which are localized in each
site of the lattice, and provides a simple framework allowing fully-analytical
developments. Analytic solutions describing the particle motion are
explicit derived, and the resulting dynamics is studied. 
}


\maketitle

\section{\label{sec:intro} Introduction}

A recent revival of interest on the old problem of electron motion
in perfect, non-dissipative, periodic lattices \cite{Bloch_States_ZFP28,Zener_BlochOsc_PRSA34}
is due to the advent of laser cooling techniques, that have opened
the possibility of realizing the interaction of atoms with perfect,
non-dissipative, \emph{optical lattices}. Optical lattices are a consequence
of the displacement of atomic levels resulting from the interaction
with light \cite{CCTHouches90,MeystreAtOpt}. A far-detuned standing
wave formed by two counter-propagating laser beams (of wave-vector
$k_{L}$) is perceived by an atom as a one-dimensional potential $V$,
whose strength varies sinusoidally in the space: \textbf{$V\propto \cos (2k_{L}x)$}
(where \textbf{$x$} is the spatial variable). Standing waves are
the model potential considered in the present work. The main source
of dissipation in such lattices is spontaneous emission, which can
be reduced in a controlled way by increasing the laser-atom detuning.
By scanning the frequency of one of the beams forming the standing
wave with respect to the other, one can generate an accelerated potential.
In the (non-inertial) frame in which the standing wave is at rest,
an inertial constant force appears, producing a {}``tilted'' lattice.
This technique can also be used to modulate (temporally) the position
or the slope of the potential. Tilted optical lattices generated in
this way have been used for recent experimental observations of Bloch
oscillations both with atoms \cite{Salomon_BlochOsc_PRL96} or Bose-Einstein
condensate \cite{Arimondo_BECBlochOsc_PRL01}. Wannier-Stark ladders
\cite{Raizen_WSOptPot_PRL96} and collective tunneling effects \cite{Kasevich_BECTiltedLattice_Science98}
have also been studied with such a system. 

A very convenient theoretical framework for studying atom dynamics
in a tilted lattice is that formed by the so-called \emph{Wannier-Stark}
(WS) states \cite{Wannier_WS_PR60}, which are the eigenstates of
a tilted lattice \cite{Korsch_LifetimeWS_PRL99,AP_WannierStark_PRA02}.
In particular, we have studied in a previous work the quantum dynamics
in a time-modulated, tilted lattice using such basis \cite{Nienhuis_CoherentDyn_PRA01,AP_WannierStark_PRA02}.
In the present paper, we present exact solutions of the equations
of motion obtained in Ref. \cite{AP_WannierStark_PRA02} that describe
the atomic center of mass motion. Sec. \ref{sec:WSBasis} briefly
reviews the main properties of WS states leading to a simple description
of Bloch oscillations; Sec. \ref{sec:ModLattice} studies the coherent
dynamics in a modulated lattice, leading to a general equation of
motion; we derive in Sec. \ref{sec:ExactSolutions} an exact solution
for the equation of motion which is studied and characterized in Sec.
\ref{sec:WavepackDyn}.

\section{\label{sec:WSBasis} The Wannier-Stark basis}

Let us first introduce natural units in which lengths are measured
in units of the lattice step $d$ ($=\lambda _{L}/2$, $\lambda _{L}$
$=2\pi /k_{L}$ being the laser wavelength), energy is measured in
units of the {}``recoil energy'' $E_{R}=\hbar ^{2}k_{L}^{2}/(2M)$
where $M$ is the atom mass, and time is measured in units of $\hbar /E_{R}$.
The Hamiltonian corresponding to a tilted lattice is then: \begin{equation}
H_{0}=\frac{P²}{2m}+V_{0}\cos (2\pi x)+Fx\label{eq:H0}\end{equation}
 where $m=\pi ^{2}/2$ is a reduced mass, $F$ is a constant force
measured in units of $E_{R}/d$, the momentum operator in real space
is $p=-i(\partial /\partial x)$, and $\hbar =1$. 

\begin{figure}
\includegraphics[height=8cm,angle=270]{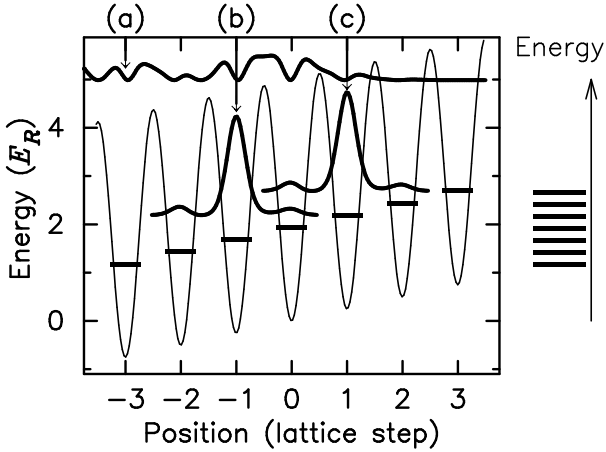}

\caption{\label{fig:WS}The tilted lattice and Wannier-Stark states. The WS
state labeled (a) is a delocalized state of the continuum. States
labeled (b) and (c) are lowest-energy states localized, resp., on
the sites $n=-1$ and $n=1$ and are identical except by a translation
of two lattice steps. The energy levels drawn at the right of the
plot show the Wannier-Stark ladder.}
\end{figure}

As shown in Fig. \ref{fig:WS}, the symmetries of the tilted lattice
suggest that the eigenenergies shall form {}``ladder'' structures
separated by the {}``Bloch frequency'' $\omega _{B}=F$ (or $\omega _{B}=Fd/\hbar $
in usual units), the so-called {}``Wannier-Stark ladders'' \cite{Wannier_WS_PR60}.
Each ladder corresponds to a family (labeled $m$) of eigenfunctions
$\varphi _{nm}(x)$ (the Wannier-Stark states) centered at the well
$n$, and thus spatially separated by an integer multiple of lattice
steps. Inside each family, WS states are invariant under a translation
by an integer number $n$ of lattice steps, provided the associated
energy is also shifted by the same number of $\omega _{B}$, i.e.
\[
\varphi _{nm}(x)=\varphi _{0m}(x-n)\]
 and \begin{equation}
E_{nm}=E_{0m}+n\omega _{B}\label{eq:EnergyTranslation}\end{equation}
 We consider a spatially limited lattice, extending over many periods
of the potential, enclosed in a large bounding box %
\footnote{In the absence of a bounding box, Wannier-Stark states are metastable
states (resonances).%
}. Although the presence of the bounding box breaks, strictly speaking,
the above symmetries, it changes only very slightly the properties
of {}``bulk'' states, and the eigenenergies and eigenstates obtained
numerically display, as long as we stay far from the bounding box,
these symmetry properties to a very good precision. The existence
of WS states has been evidenced experimentally in 1988 in a semi-conductor
superlattice \cite{Mendez_WSSuperlattice_PRL88,Voisin_WSSuperlattice_PRL88},
and 1996 with cold atoms in an optical lattice \cite{Raizen_WSOptPot_PRL96}. 

We shall consider here parameters such that WS states are strongly
localized inside a given lattice well, and that the atom dynamics
can be described to a good accuracy by the lowest-energy WS state
in each well (as the states noted (b) or (c) in Fig. \ref{fig:WS})
\footnote{\label{note:LandauZener}This is equivalent to neglect Landau-Zener
inter-band transitions in the usual Bloch-function approach.%
}. We thus drop the family index $m$, and note the WS states by $\varphi _{n}(x)$,
with associated energy $E_{n}=n\omega _{B}$ (setting $E_{0}=0$). 

The system dynamics is described by projecting the atomic wave function
on the WS basis: \begin{equation}
\Psi (x,t)=\sum _{n}c_{n}(t)\varphi _{n}(x)\label{eq:PsiWS}\end{equation}
 with $c_{n}(t)=c_{n}(0)e^{-in\omega _{B}t}$. Dynamical observables
can be easily calculated. For instance, the mean value of the atomic
position operator, $\langle x\rangle _{t}=\left\langle \Psi \left|X\right|\Psi \right\rangle $
is: \begin{equation}
\langle x\rangle _{t}=\sum _{n}X_{nn}\left|c_{n}(t)\right|^{2}+\sum _{n<m}\left(X_{nm}c_{n}^{\ast }(t)c_{m}(t)+\textrm{c.c.}\right)\label{eq:XmeanGen}\end{equation}
 where $X_{nm}\equiv \left\langle \varphi _{n}\left|X\right|\varphi _{m}\right\rangle $
are constants depending only on the parameters $V_{0}$ and $F$ of
the tilted lattice. As we are considering WS states localized in the
wells, we can keep only nearest-neighbors contributions ($\left|n-m\right|\leq 1)$
and derive a simplified expression: \begin{equation}
\langle x\rangle _{t}=\overline{x}+X_{1}\left(\sum _{n}c_{n}(0)^{\ast }c_{n+1}(0)e^{-i\omega _{B}t}+\textrm{c.c.}\right)\label{eq:Xmean}\end{equation}
 where $\overline{x}=\sum X_{nn}\left|c_{n}(0)\right|^{2}$ is the
mean position of the wave packet and $X_{1}\equiv X_{n\left(n+1\right)}(=X_{01}=X_{0-1}=X_{n\left(n-1\right)})$
is independent of $n$ %
\footnote{\label{note:Identities}By virtue of the translational properties
of the WS states, $X_{nn+p}$ ($p\neq 0$) does not depend on $n$:
$X_{nn+p}=\int \varphi _{n}^{\ast }(x)x\varphi _{n+p}(x)dx=\int \varphi _{0}^{\ast }(x-n)x\varphi _{p}(x-n)dx=\int \varphi _{0}^{\ast }(x^{\prime })(x^{\prime }+n)\varphi _{p}(x^{\prime })dx^{\prime }=X_{0p}\equiv X_{p}$,
where we used the orthogonality of the WS states. The same kind of
calculation leads to $X_{nn}=X_{00}+n$.%
}. Eq. (\ref{eq:Xmean}) describes Bloch oscillations, in a simpler
and more intuitive way than the usual Bloch-function approach \cite{Zener_BlochOsc_PRSA34,Holthaus_BlochOsc_JOBQSO}.
The amplitude of the Bloch oscillation is proportional to $X_{1}$,
and grows with the overlap between neighbors WS states, i.e. it increases
as the slope $F$ and lattice depth $V_{0}$ decrease. The physical
origin of the Bloch oscillations appears very clearly as an interference
effect between neighbor sites, since $c_{n}^{\ast }c_{n+1}$ is the
quantum coherence between the sites $n$ and $n+1$ \cite{Nienhuis_CoherentDyn_PRA01}.
Bloch frequency is seen to be just the fundamental \emph{Bohr} frequency
of the system.

\section{\label{sec:ModLattice} Coherent dynamics in a modulated lattice}

We now investigate the case in which the system is {}``forced''
with a frequency close to its natural frequency $\omega _{B}$. We
therefore add a sinusoidal component of frequency $\omega $ {[}i.e.,
a term of the form $F_{0}\sin (\omega t)x${]} to the Hamiltonian
Eq. (\ref{eq:H0}), with $\omega =\omega _{B}+\delta $ and $\left|\delta \right|\ll \omega ,\omega _{B}.$
The WS basis again leads to an analytical approach \cite{AP_WannierStark_PRA02}.
We limit ourselves to modulations that are smooth enough in order
to avoid transitions to \emph{delocalized} or to localized \emph{excited}
WS states, ensuring that the decomposition over the lowest-energy
state of each well {[}Eq. (\ref{eq:PsiWS}){]} remains valid for all
times. 

The coefficients $c_{n}(t)$ can then be obtained by reporting Eq.
(\ref{eq:PsiWS}) into the Schrödinger equation \begin{equation}
i\frac{\partial \Psi (x,t)}{\partial t}=\left[\frac{P²}{2m}+V_{0}\cos (2\pi x)+Fx-F_{0}x\sin (\omega t)\right].\label{eq:ModulatedH}\end{equation}
 which produces the following set of coupled differential equations
for the $c_{n}(t)$: \[
\dot{c}_{n}(t)=-in\omega _{B}c_{n}(t)+iF_{0}\sin (\omega t)\sum _{m}X_{nm}c_{m}(t)\]
 where $\dot{c}_{n}\equiv dc_{n}/dt$. Neglecting temporarily the
coupling between different WS states, (i.e putting $X_{nm}=X_{nn}\delta _{mn}$),
the amplitudes are obtained simply as $c_{n}=\exp \left[i\phi _{n}(t)\right]$,
with the time-dependent phase \begin{equation}
\phi _{n}(t)=-n\omega _{B}t-\left(\frac{F_{0}}{\omega }\right)\left(X_{00}+n\right)\cos (\omega t)\label{eq:phase}\end{equation}
 where we used the identity $X_{nn}=X_{00}+n$ \ref{note:Identities}.
Writing \begin{equation}
c_{n}(t)\equiv d_{n}(t)e^{i\phi _{n}(t)},\label{eq:cndn}\end{equation}
 the new amplitudes $d_{n}$ are governed by: \begin{equation}
\dot{d}_{n}=iF_{0}\sum _{m\neq n}X_{nm}d_{m}(t)\exp \left\{ i\left[\phi _{m}(t)-\phi _{n}(t)\right]\right\} \sin (\omega t).\label{eq:dn0}\end{equation}
 After Eq. (\ref{eq:phase}), the phase difference $\phi _{m}(t)-\phi _{n}(t)$
is: \[
\phi _{m}(t)-\phi _{n}(t)=(n-m)\left[\omega _{B}t+\frac{F_{0}}{\omega }\cos (\omega t)\right]\]
 where we used Eq. (\ref{eq:EnergyTranslation}). Eq. (\ref{eq:dn0})
can be recast as: 
\begin{eqnarray}
\dot{d_{n}} & = & iF_{0}\sum _{p\neq 0}X_{p}d_{n+p}\left[e^{-ip\omega _{B}t}e^{-ip\left(F_{0}/\omega \right)\cos \left(\omega t\right)}\right]\sin (\omega t)\nonumber \\
 & = & \frac{F_{0}}{2}\sum _{p\neq 0}X_{p}d_{n+p}\sum _{\ell }\left(-i\right)^{\ell }J_{\ell }
\left(p\frac{F_{0}}{\omega }\right) \nonumber \\
& \times & \left[ \exp \left\{ i\left[(\ell +1)\omega -p\omega _{B}\right]t\right\} \right. \nonumber \\
& - & \left. \exp \left\{ i\left[(\ell -1)\omega -p\omega _{B}\right]t\right\} \right]\label{eq:dn1}
\end{eqnarray}
 where $X_{p}\equiv X_{n(n+p)}$ $(p\neq 0)$ \ref{note:Identities}.
$J_{n}(x)$ is the Bessel function of the first kind, and we have
used the well-known formula: \begin{equation}
e^{-iz\cos (\omega t)}=\sum _{\ell =-\infty }^{+\infty }J_{\ell }(z)(-i)^{\ell }e^{i\ell \omega t}.\label{eq:BesselGenerator}\end{equation}

The coupling parameters $X_{p}$ rapidly shrinks to zero as $p$ increases
and we can keep, to a good accuracy, only the contribution of the
neighbor site ($p=1$) (for instance, $X_{1}=5\times 10^{-2}$ , $X_{2}=8\times 10^{-4}$
for $V_{0}=4.5$ and $F=0.5$). Moreover, the sum over the harmonics
of $\omega $ (i.e. $\ell $) is also limited to a few terms close
to $\ell =0$ {[}typically, $\ell _{max}\sim pF_{0}/\omega \sim O(1)${]}
\footnote{\label{note:Besselvalue}The Bessel function value $J_{n}(x)$ becomes
small for $\left|x\right|\gtrsim \left|n\right|$.%
}. Finally, we keep only the slowly varying terms in Eq. (\ref{eq:dn1})
: since $\delta =\omega -\omega _{B}$ is assumed small ($\left|\delta \right|\ll \omega ,\omega _{B})$,
on can retain, to a very good accuracy, only the terms which oscillate
as $\exp (\pm i\delta t)$ (the fast oscillations give a vanishing
contribution on the average). To the leading order, this gives: \begin{equation}
\dot{d}_{n}(t)=\Omega _{1}\left[d_{n+1}e^{i\delta t}-d_{n-1}e^{-i\delta t}\right]\label{eq:dnsimple}\end{equation}
 where \begin{equation}
\Omega _{1}=\frac{F_{0}X_{1}}{2}\left[J_{0}\left(\frac{F_{0}}{\omega }\right)+J_{2}\left(\frac{F_{0}}{\omega }\right)\right]=\omega X_{1}J_{1}\left(\frac{F_{0}}{\omega }\right).\label{eq:RabiFreq}\end{equation}
 Equation (\ref{eq:dnsimple}) is similar to a {}``dipole coupling''
between sites $n$ and $n\pm 1$, where $\Omega _{1}$ plays the role
of a Rabi frequency %
\footnote{Eq. (\ref{eq:RabiFreq}) predicts that the coupling vanishes if $F_{0}/\omega $
coincides with a zero of $J_{1}(x)$. However, the first root of this
function arises for $x\simeq 3.8$, that is, for $F_{0}\sim 3.8\omega =3.8F$,
which cannot be considered as a smooth modulation, falling outside
the range of validity of the approximations leading to Eq. (\ref{eq:RabiFreq}).%
}. In the next section, we will find an exact solution for the above
equation.

\section{\label{sec:ExactSolutions}Exact solution}

It is interesting to consider the complex amplitudes $d(k,t)$ defined
in the reciprocal space: \begin{eqnarray}
d(k,t) & = & \sum _{n=-\infty }^{\infty }d_{n}(t)e^{ink}\label{eq:TF1}\\
d_{n}(t) & = & \frac{1}{2\pi }\int _{-\pi }^{\pi }d(k,t)e^{-ink}dk.\label{eq:TF2}
\end{eqnarray}
 In these expressions, $d_{n}(t)$ are the Fourier coefficients of
the continuous and periodic function (period $2\pi )$ $d(k,t),$
and $k$ plays the role of a quasi-momentum. Eq. (\ref{eq:dnsimple})
can easily be translated to the $k$-space: \begin{equation}
\dot{d}(k,t)=2i\Omega _{1}\sin (\delta t-k)d(k,t).\label{eq:dk}\end{equation}
 The quasi-momentum distribution is obtained after a straightforward
integration: 
\begin{eqnarray}
d(k,t)& = &d(k,0)\exp \left\{ \frac{2i\Omega _{1}}{\delta }\left[1-\cos (\delta t)\right]\cos (k)\right\} \nonumber \\
& \times & \exp \left\{ \frac{2i\Omega _{1}}{\delta }\sin (\delta t)\sin (-k)\right\} \label{eq:dkt}
\end{eqnarray}
 and, using Eq. (\ref{eq:BesselGenerator}) and the equivalent relation
$e^{iz\sin (-k)}=\sum _{\ell }J_{\ell }(z)e^{-i\ell k}$ : 
\begin{eqnarray}
d(k,t)&=&d(k,0)\sum _{p,\ell }J_{p}\left(\frac{2\Omega _{1}}{\delta }\left[1-\cos (\delta t)\right]\right)\nonumber \\
&\times& J_{\ell }\left(\frac{2\Omega _{1}}{\delta }\sin (\delta t)\right)
e^{ip(k+\frac{\pi }{2})}e^{-i\ell k}.
\end{eqnarray}
 Setting $\ell =p+q$: 
\begin{eqnarray}
d(k,t)&=&d(k,0)\sum _{p,q}J_{p}\left(\frac{2\Omega _{1}}{\delta }\left[1-\cos (\delta t)\right]\right)\nonumber \\
&\times& J_{p+q}\left(\frac{2\Omega _{1}}{\delta }\sin (\delta t)\right)e^{ip\frac{\pi }{2}}e^{-iqk};
\label{eq:dkt0}
\end{eqnarray}
 and using the Bessel addition theorem (see Appendix \ref{sec:Bessel-sum}),
we transform the above equation into: \[
d(k,t)=d(k,0)\sum _{q}J_{q}(Q)e^{-iqk}e^{iq\delta t/2}\]
 where \[
Q(t)\equiv \frac{4\Omega _{1}}{\delta }\sin (\frac{\delta t}{2}).\]
 It is now straightforward to obtain the coefficients $d_{n}(t)$
by applying Eq. (\ref{eq:TF2}): \begin{eqnarray}
d_{n}(t) & = & \sum _{q}d_{n+q}(0)J_{q}(Q)e^{iq\delta t/2}\label{eq:dnt}\\
 & = & \sum _{m}d_{m}(0)J_{m-n}(Q)e^{i(m-n)\delta t/2}\nonumber 
\end{eqnarray}
 This remarkably simple result shows that amplitude of probability
of finding the particle in the state $n$ at time $t$ is simply the
sum of the contributions of the initial populated states $m,$ $d_{m}(0)$,
with a time dependent weight $J_{m-n}\left(Q\right)e^{i(m-n)\delta t/2}$,
that depends on the distance $n-m$ between the sites. The presence
of the evolving phase term $e^{i(m-n)\delta t/2}$ evidences the quantum-coherent
nature of the dynamics. 

\begin{figure}
\includegraphics[height=8cm,angle=270]{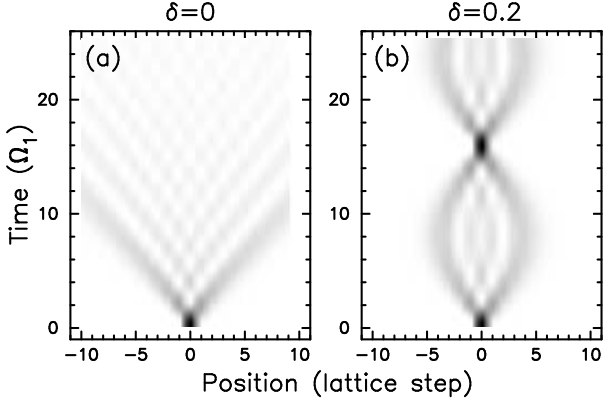}

\caption{\label{fig:Dynamique}The gray-level plots show the temporal evolution
of the probability of presence of the atom, initially localized in
the site $m=0$. Plot (a) shows the resonant behavior: the wave packet
spreads linearly with time. Plot (b) displays the non-resonant {}``breathing''
behavior. ($\delta $ is indicated in units of $4\Omega _{1}$).}
\end{figure}

The most relevant contributions to the above summation are due to
sites such that $\left|n-m\right|\lesssim \left|Q(t)\right|$. Roughly
speaking, the dynamics of site $n$ is correlated to the sites situated
inside a range $\Delta n\approx \left|Q(t)\right|$ around $n$, which,
out for resonance, is roughly $\left|4\Omega _{1}/\delta \right|$.
Close to the resonance, this range is $\Delta n\approx 2\left|\Omega _{1}\right|t$,
and the dynamics on site $n$ will {}``mix'' contributions of larger
and larger numbers of sites as time increases. To give a simple example,
consider the case $d_{m}(0)=\delta _{m,0}$. One then finds ($\delta \neq 0$):
\[
\left|d_{n}(t)\right|^{2}=\left[J_{n}(Q)\right]^{2}\]
 which shows that the $n^{\textrm{th}}$ site is {}``filled'' and
{}``emptied'' in a periodic way with a period \textbf{$2\pi /\delta $}.
This {}``breathing'' behavior is shown in Fig \ref{fig:Dynamique}. 

\section{\label{sec:WavepackDyn}Wave packet dynamics}

The dynamics of a wave packet can be characterized by calculating
mean values of observables. This can be done analytically in the present
framework. One can easily deduce from Eq. (\ref{eq:XmeanGen}) the
mean position: \begin{equation}
\langle x\rangle _{t}=\sum _{n}X_{nn}\left|d_{n}(t)\right|^{2}+X_{1}\sum _{n}\left(d_{n}(t)^{\ast }d_{n+1}(t)e^{i\phi (t)}+c.c.\right)\label{eq:xmoyen}\end{equation}
 where $\phi (t)=\phi _{1}(t)-\phi _{0}(t)=-\omega _{B}t-(F/\omega )\cos (\omega t)$.
Using the relation $X_{nn}=X_{00}+n\equiv X_{0}+n$, a straightforward
although somewhat lengthy calculation (see Appendix \ref{sec:Meanvalues})
produces: 
\begin{eqnarray}
\langle x\rangle _{t}&=&\langle x\rangle _{t=0}+\left\{ \left[-\frac{Q(t)}{2}e^{-i\delta t/2}+X_{1}\left(e^{i\phi (t)}-e^{i\phi (0)}\right)\right] \right. \nonumber \\
& \times & \left. \sum _{n}d_{n}^{\ast }(0)d_{n+1}(0)+c.c\right\} .
\label{eq:xmean-exact}
\end{eqnarray}
 The mean position will evolve with time only if the initial packet
displays site-to-site quantum coherence, i.e. if $\sum _{n}d_{n}^{\ast }(0)d_{n+1}(0)\neq 0$
(this is also a condition for observing Bloch oscillations, as pointed
out in Sec. \ref{sec:WSBasis}). The evolution in time may display
both a fast oscillation at the frequencies $\omega _{B}$ and $\omega $
(and its harmonics) and of a slow oscillation at the frequency $\delta $.
Note that the weight ratio between the slow and the fast component
is roughly $\Omega _{1}/(\delta X_{1})$: the slow oscillation is
dominant close to resonance. 

The meaning of the above result can be more clearly appreciate by
considering the simple case in which the initial wave packet spreads
over $N$ sites and presents a constant phase difference $e^{ik_{0}}$
from site to site: 
\begin{equation}
d_{n}(0)=\frac{e^{ik_{0}n}}{\sqrt{N}}.\label{eq:WSWave}
\end{equation}
Substitution into Eq. (\ref{eq:xmean-exact}) leads to: 
\begin{eqnarray}
\langle x\rangle _{t} &=&\langle x\rangle _{t=0}+\left\{ -Q(t)\cos (k_{0}-\delta t/2) \right. \nonumber \\
&+& \left. 2X_{1}\cos \left[k_{0}-\omega _{B}t-(F_{0}/\omega )\cos (\omega t)\right] \right. \nonumber \\
&-& \left. 2X_{1}\cos \left(k_{0}-F_{0}/\omega \right)\right\} 
\end{eqnarray}
 If $\delta =0$: 
\begin{eqnarray}
\langle x\rangle _{t}-\langle x\rangle _{t=0} &=& -2\Omega _{1}\cos (k_{0})t \nonumber \\
&+& 2X_{1}\left(\cos \left[k_{0}-\omega _{B}t-(F_{0}/\omega _{B})\cos (\omega _{B}t)\right] \right. \nonumber \\
&-& \left. \cos \left(k_{0}-F_{0}/\omega _{B}\right)\right)=v_{g}t+f(t)
\label{eq:vg}
\end{eqnarray}
 where $v_{g}=-2\Omega _{1}\cos (k_{0})$ is the group-velocity of
the wave packet, and $f(t)$ a purely oscillating function. This result
shows that the main features of the resonant dynamics are determined
by $k_{0}$. If $k_{0}=\pm \pi /2$ (phase-quadrature from site to
site), $v_{g}=0$ and there is no global motion (but there is an oscillation
and a spreading of the wave packet). If $k_{0}=\pi $, $v_{g}=2\Omega _{1}$
and the global motion is a climb up along the slope of the potential
with the maximum speed $2\Omega _{1}$. In the case $k_{0}=0$, where
$v_{g}=-2\Omega _{1}$ the atom climbs down the slope of the potential
with a constant maximum speed: there is \emph{coherent transfer of
energy} from the modulation to atom, thanks to the particular phase
relation between neighbor sites. Contrary to the motion of a classical
particle, the speed $\left|v_{g}\right|$ is independent of the sense
of the motion: the wave packet climbs the slope up or down at the
same speed. 

Let us finally note that the spreading of the wave packet $\left\langle x^{2}\right\rangle _{t}$
can also be obtained analytically. However, this leads to heavier
formulas, that the interested reader can find in the Appendix \ref{sec:Meanvalues}. 

Before closing this section, let us note that other resonant behaviors
can be observed for $\omega =q\omega _{B}$ ($q$ integer). From the
general expression of Eq. (\ref{eq:dn1}), one finds for a coupling
between sites of the form ($n\rightarrow n\pm q$), for which the
same techniques can be applied. For example, resonant interaction
with $\omega =2\omega _{B}$, leads to: \begin{equation}
\dot{d}_{n}(t)=\Omega _{2}\left[d_{n+2}-d_{n-2}\right]\label{eq:DynNextNeig2}\end{equation}
 with $\Omega _{2}=\omega X_{2}J_{1}(F_{0}/\omega )$. Numerical calculations
show that $X_{2}\ll X_{1}$. Around this resonance, interesting kinds
of coherent dynamics can be observed \cite{AP_WannierStark_PRA02}.

\section{conclusion}

We have developed an analytical approach to the dynamics of a wave
packet in static and time-modulated tilted lattices, and provided
exact solutions describing the atomic motion. It is worth noting that
the present description is, in its principle, independent of the shape
of the lattice, provided it presents localized states. It is therefore
generalizable to non-sinusoidal lattices. 

The major experimental difficulty for observing the effects described
in this paper is the creation of the initial atomic coherence. There
are various solutions for that. One is to cool the atoms to sub-recoil
temperatures, so that their de Broglie wavelength is of the order
of a few lattice steps (5-10 lattice steps is typical) \cite{Salomon_BlochOsc_PRL96};
the light potential is then turned on adiabatically. Another way of
creating spatial coherence is to start with a Bose-Einstein condensate,
whose spatial coherence length is up to hundreds lattice sites \cite{Kasevich_BECTiltedLattice_Science98,Arimondo_BECBlochOsc_PRL01}.
Note that in the latter case the system obeys a nonlinear Schr\"{o}dinger
equation, but the present approach can still be generalized to such
case \cite{AP_ChaosBEC_ARXIV_03}. 

The detection of the coherent dynamics described above is simpler
in the \emph{momentum} space, using velocity-sensitive Raman stimulated
transitions \cite{Chu_RamanVSel_PRL91,Salomon_BlochOsc_PRL96,AP_RamanSpectro_PRA02}.
The coherent motion of atoms, e.g. when they climb the slope of the
modulated potential (Sec. \ref{sec:ModLattice}), correspond to a
speed of about a recoil velocity, i.e 3 mm/s for cesium, which can
be easily detectable by Raman stimulated spectroscopy. It would be
nice to observe the coherent motion also in the real space. This seems
to be very difficult, because the spatial amplitude of the motion
is very small, of the order of a few lattice steps, as compared to
the atomic cloud which extends over hundreds of lattice wells (for
atoms cooled in a Magneto-Optical trap). The coherent dynamics is
thus detectable in momentum space, but very hard to see in the real
space, at least under usual experimental conditions. 

\acknowledgement{
The authors are grateful to S. Bielawski for fruitful discussions.
This work is partially supported by a contract {}``ACI Photonique''
of the Ministère de la Recherche. Laboratoire de Physique des Lasers,
Atomes et Mol\'{e}cules (PhLAM) is Unit\'{e} Mixte de Recherche
UMR 8523 du CNRS et de l'Universit\'{e} des Sciences et Technologies
de Lille. Centre d'Etudes et Recherches Lasers et Applications (CERLA)
is Fédération de Recherche FR 2416 du CNRS. 
}
\appendix

\section{Bessel addition theorem\label{sec:Bessel-sum}}

The addition theorem for Bessel functions of integer order states
that: \begin{equation}
\sum _{p}J_{p}(r_{0})e^{-ip\theta _{0}}J_{p+q}(r_{1})e^{i(p+q)\theta _{1}}=J_{q}(Q)e^{iq\Theta }\label{eq:adtheo0}\end{equation}
 where $\overrightarrow{Q}=\overrightarrow{r}_{1}-\overrightarrow{r}_{0}\equiv R$
$e^{i\Theta }$, and $\overrightarrow{r_{1}}=r_{1}e^{i\theta _{1}}$,
$\overrightarrow{r_{0}}=r_{0}e^{i\theta _{0}}$. For Eq. (\ref{eq:dkt0}),
one sets $r_{0}=\frac{2\Omega _{1}}{\delta }\left[1-\cos (\delta t)\right],$
$\theta _{0}=-\pi /2$ $r_{1}=\frac{2\Omega _{1}}{\delta }\sin (\delta t)$,
$\theta _{1}=0$, and finds quite straightforwardly\begin{eqnarray*}
Q & = & \sqrt{r_{1}^{2}+r_{0}^{2}}\\
 & = & \frac{4\Omega _{1}}{\delta }\sin \frac{\delta t}{2}
\end{eqnarray*}
 and \[
\Theta =\frac{\delta t}{2}\]
 Eq. (\ref{eq:dn1}) takes the following form: 
\begin{eqnarray*}
d(k,t) & = & d(k,0)\sum _{p,q}J_{p}\left(\frac{2\Omega _{1}}{\delta }\left[1-\cos (\delta t)\right]\right)\nonumber \\
& \times & J_{p+q}\left(\frac{2\Omega _{1}}{\delta }\sin (\delta t)\right)e^{ip\frac{\pi }{2}}e^{-iqk}\\
 & = & d(k,0)\sum _{q}J_{q}\left(Q\right)e^{-iqk}e^{iq\frac{\delta t}{2}}
\end{eqnarray*}

\section{Mean values\label{sec:Meanvalues}}

Consider first the first term on the R.H.S of Eq. (\ref{eq:xmoyen}):
\begin{eqnarray}
\left\langle x\right\rangle _{t}^{(1)}&=&\sum _{n}(X_{0}+n)\sum _{m}J_{m}(Q)d_{n+m}^{\ast }(0)e^{-im\delta t/2}\nonumber \\
& \times & \sum _{\ell }J_{\ell }(Q)d_{n+\ell }(0)e^{i\ell \delta t/2}.
\end{eqnarray}
 Changing variables so that $q=m-\ell $, $p=n+\ell $ leads to: \begin{eqnarray*}
\left\langle x\right\rangle _{t}^{(1)} & = & \sum _{n,q,\ell }\left(X_{0}+n\right)d_{n+\ell }(0)d_{n+\ell +q}^{\ast }(0)J_{\ell }J_{\ell +q}e^{-iq\delta t/2}\\
 & = & \sum _{p,q,\ell }\left(X_{0}+p-\ell \right)d_{p}(0)d_{p+q}^{\ast }(0)J_{\ell }J_{\ell +q}e^{-iq\delta t/2}
\end{eqnarray*}
 where $J_{n}\equiv J_{n}\left(Q\right)$, $Q=\frac{4\Omega _{1}}{\delta }\sin (\frac{\delta t}{2})$.
Eq. \ref{eq:adtheo0} implies here: \begin{equation}
\sum _{\ell }J_{\ell }(Q)J_{\ell +q}(Q)=J_{q}(0)=\delta _{q,0}.\label{eq:bessel-sum}\end{equation}
 and using also the Bessel recurrence: \begin{equation}
\ell J_{\ell }(Q)=\frac{Q}{2}\left(J_{\ell +1}(Q)+J_{\ell -1}(Q)\right)\label{eq:bessel1}\end{equation}
 one gets then \begin{eqnarray*}
\left\langle x\right\rangle _{t}^{(1)} & = & \sum _{p}\left(X_{0}+p\right)\left|d_{p}(0)\right|^{2}-\\
 &  & \left(\frac{Q}{2}e^{-i\delta t/2}\sum _{p}d_{p}(0)d_{p+1}^{\ast }(0)+c.c.\right).
\end{eqnarray*}
sub It is simpler to show, by the same kind of reasoning, that the
second term in Eq. (\ref{eq:xmoyen}) is \begin{eqnarray*}
\left\langle x\right\rangle _{t}^{(2)} & = & X_{1}\sum _{p}\left(d_{p}(t)^{\ast }d_{p+1}(t)e^{i\phi (t)}+c.c.\right)\\
 & = & X_{1}\sum _{p}\left(d_{p}(0)^{\ast }d_{p+1}(0)e^{i\phi (t)}+c.c.\right),
\end{eqnarray*}
 so that the mean is: 
\begin{eqnarray*}
\left\langle x\right\rangle _{t} = \left\langle x\right\rangle _{t=0} &+&
\left\{ \left[-\frac{Q}{2}e^{-i\delta t/2}+X_{1}\left(e^{i\phi (t)}-e^{i\phi (0)}\right)\right] \right.  \nonumber \\
& \times & \left. \left(\sum _{p}d_{p}(0)^{\ast }d_{p+1}(0)\right)+c.c.\right\} 
\end{eqnarray*}
 with 
\begin{eqnarray*}
\left\langle x\right\rangle _{t=0}&=&\sum _{p}\left(X_{0}+p\right)\left|d_{p}(0)\right|^{2}\nonumber \\
&+& X_{1}e^{i\phi (0)}\sum _{p}\left(d_{p}(0)^{\ast }d_{p+1}(0)+c.c.\right)
\end{eqnarray*}. 

One can calculate, by the same kind of technique, the spreading of
the wave packet: 
\begin{eqnarray}
\left\langle x^{2}\right\rangle _{t} & = & \left\langle x^{2}\right\rangle _{t=0}-
 Q\sum _{p}\left\{ \left[\left(X_{0}+p+\frac{1}{2}\right)e^{-i\delta t/2} \right. \right. \nonumber \\
&+& \left. \left. X_{1}^{(2)}\left(e^{i\phi (t)}-e^{i\phi (0)}\right)\right]
d_{p}^{\ast }(0)d_{p+1}(0)+c.c.\right\} \nonumber \\
&+&  \frac{Q^{2}}{4}\sum _{p}\left[\left(d_{p}^{\ast }(0)d_{p+2}(0)e^{-i\delta t}+\left|d_{p}(0)\right|^{2}+c.c.\right)\right]
\label{eq:X2mean}
\end{eqnarray}
 where $X_{1}^{(2)}=\int \varphi _{0}^{\ast }(x)x^{2}\varphi _{1}(x)$, and 
\begin{eqnarray*}
\left\langle x^{2}\right\rangle _{t=0}&=&\sum _{p}\left(X_{0}^{(2)}+2pX_{0}+p^{2}\right)\left|d_{p}(0)\right|^{2}\nonumber \\
&+& X_{1}^{(2)}e^{i\phi (0)}\sum _{p}\left(d_{p}(0)^{\ast }d_{p+1}(0)+c.c.\right)
\end{eqnarray*}
with $X_{0}^{(2)}=\int \varphi _{0}^{\ast }(x)x^{2}\varphi _{0}(x)=\int \varphi _{n}^{\ast }(x)x^{2}\varphi _{n}(x)-2nX_{0}-n^{2}$. 

%\bibliographystyle{unsrt}
%\bibliography{/home/jcg/papers/ArtDataBase,/home/jcg/papers/Books}

\end{document}